# Browserbite: Cross-Browser Testing via Image Processing

Tõnis Saar[1], Marlon Dumas[1,2]*, Marti Kaljuve[1] and Nataliia Semenenko[1,2]

[1.] *STACC, Estonia*

[2.] *University of Tartu, Estonia*

## ABSTRACT

Cross-browser compatibility testing is concerned with identifying perceptible differences in the way a Web page is rendered across different browsers or configurations thereof. Existing automated cross-browser compatibility testing methods are generally based on Document Object Model (DOM) analysis, or in some cases, a combination of DOM analysis with screenshot capture and image processing. DOM analysis however may miss incompatibilities that arise not during DOM construction, but rather during rendering. Conversely, DOM analysis produces false alarms because different DOMs may lead to identical or sufficiently similar renderings. This paper presents a novel method for cross-browser testing based purely on image processing. The method relies on image segmentation to extract "regions" from a Web page and computer vision techniques to extract a set of characteristic features from each region. Regions extracted from a screenshot taken on a baseline browser are compared against regions extracted from the browser under test based on characteristic features. A machine learning classifier is used to determine if differences between two matched regions should be classified as an incompatibility. An evaluation involving 140 pages shows that the proposed method achieves an F-score exceeding 90%, outperforming a state-of-the-art cross-browser testing tool based on DOM analysis.

## 1. INTRODUCTION

Web pages are often rendered differently across multiple browsers and platforms. When these differences affect the end user experience, they are called *cross-browser incompatibilities*. Incompatibilities may take the form of invisible or overflowing text, distorted fonts or missing buttons, for example. Cross-browser compatibility testing is concerned with the identification of such defects.

Figure 1 shows a real-life example of a cross browser incompatibility. The example Web page has a distorted footer menu in Internet Explorer 9 (IE9). This cross-browser incompatibility is caused by a typographical error in the end tag of an anchor – instead of the proper end tag </a>, the element is closed with <a/>, which some browsers interpret as a new anchor element. In IE9, the improper end tag causes the next block-level element to also be wrapped in an anchor tag, resulting in an improper layout. In most other browsers, including Chrome, the end tag does not extend into the adjacent element, so the layout is unaffected. This type of error handling demonstrates inconsistent behaviour between browsers.

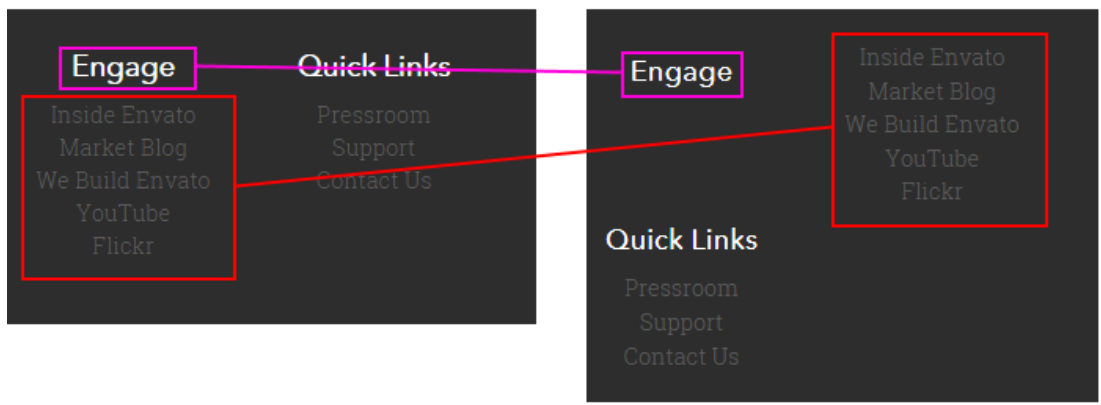

Figure 1. Footer menus on Google Chrome (left-correct) IE9 (right-defected) [1]

The problem of cross-browser testing is exacerbated by the co-existence of a wide range of browsers, browser settings and underlying platforms. Herein, we use the term *browser configuration* to refer to a browser with certain settings, running on a given operating system and device type (e.g. Firefox on Windows 7 with default settings). Manual cross-browser testing requires testers to open and inspect Web pages in each of a number of browser configurations. This task is laborious and eye straining. Automatic screen capture tools [2]-[3] reduce some of the manual labour by automating the process of opening the Web page in each browser configuration and taking a screenshot of the Web page rendering. However, they still leave the inspection of screenshots to the tester.

A naïve approach to automate the inspection step of cross-browser testing is to subtract pixel by pixel the screenshot of the Web page rendering on a *browser-under-test*, against the screenshot of a *baseline browser configuration* that has been manually validated (cf. Figure 2). In this approach, pixels with intensity differences are highlighted so that the tester can determine if a given difference constitutes an incompatibility. This naïve approach however produces a large number of false positives. Indeed, pixel-level differences can be caused by minute misalignments or differences in pixel intensity levels, which are not perceptible by end users and thus do not affect the user experience. For example, Figure 2 shows the output of a pixel-by-pixel comparison of a Chrome rendering against an Internet Explorer 10 rendering of the same Web page. The difference image (rightmost) displays a large number of pixel level differences, even though these two renderings appear visually identical, demonstrating the limits of pixel level comparison. Additionally, multiple browsers may have different viewport and element sizes, making pixel-by-pixel comparison even more ineffective.

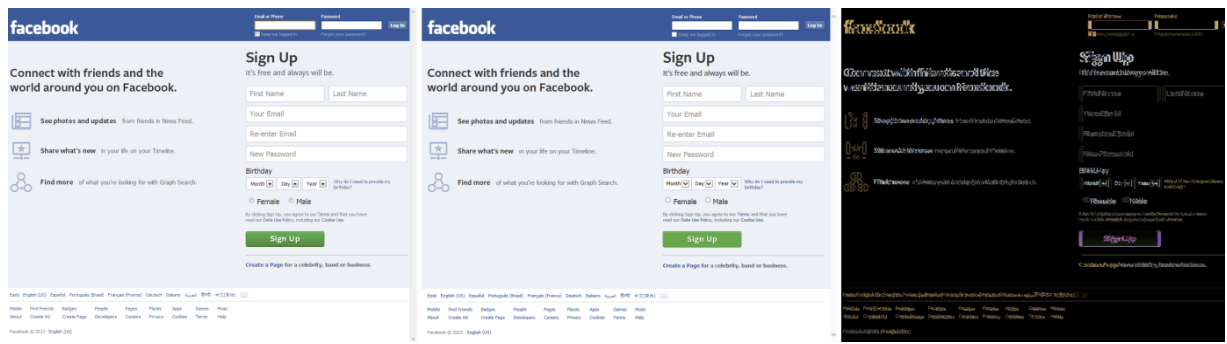

Figure 2. Google Chrome (left), Internet Explorer 10 (centre), resulting difference image (right) (www.facebook.com)

To overcome these limitations a higher abstraction level is needed. Prior to comparison, individual pixels need be aggregated into larger elements such as images, text areas, buttons and other form elements. If a Web page element is missing or distorted in the rendering of a browser-under-test, this is more likely to be an incompatibility than a pixel-level difference.

The bulk of cross-browser compatibility testing methods try to find incompatibilities based on an analysis of the DOM (Document Object Model) produced during Web page rendering [4],[2]. In such DOM-based approaches, a DOM object is generated in multiple browser configurations. The resulting DOM objects are compared and any significant difference is highlighted as a potential incompatibility – where the notion of "significant DOM difference" depends on the specific DOM-based testing method. Browsers however tend to differ considerably in terms of their DOMs. For example, Figure 3 illustrates the computed DOM parameters of a Web page [5] in Chrome and Internet Explorer 11. There is a matching DOM node that has different calculated DOM parameters across these browsers. Specifically, one of the Web page's DOM elements has size 368.688 x 15 in Chrome and 298.43 x 16 in Internet Explorer 11. This example illustrates why a direct comparison between DOM nodes is fundamentally prone to false positives. Additionally, even a perfect matching of DOMs across multiple browsers does not guarantee a similar visual rendering. This gap constitutes a source of false negatives. After more than a decade of automated cross-browser testing, it has become evident that the only way to tackle this highly persistent problem is via visual comparison of Web page renderings.

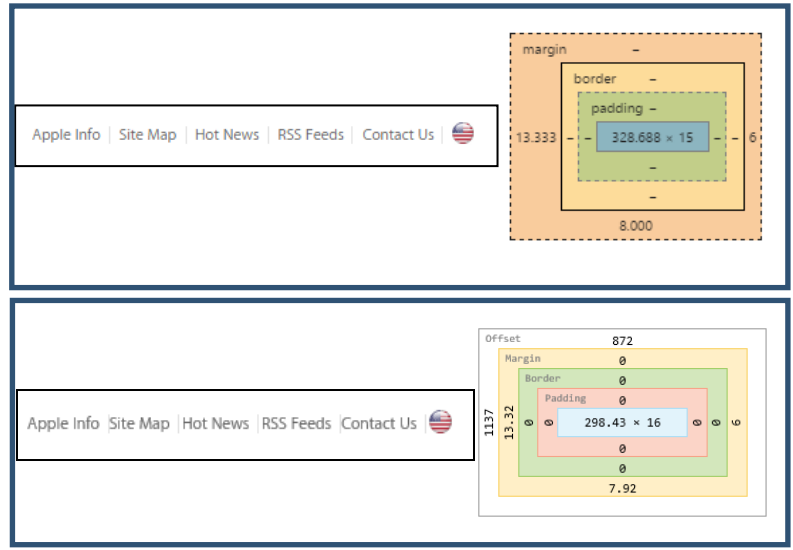

Figure 3. Comparison of DOM parameters; Google Chrome (top) vs. Internet Explorer 11 (below)

The difficulty of automated cross-browser compatibility testing is compounded by the fact that the definition of what constitutes an incompatibility is fundamentally subjective, often depending on the specific user profile. Professional Web designers for example are likely to have higher sensitivity towards cross browser differences compared to front-end developers or end users. For example, large font differences may not be considered incompatibilities by developers, whereas Web designers may have very specific expectations regarding the appearance of a given font.

In light of the above, we propose a novel method to automated cross-browser incompatibility testing, which mirrors the way cross-browser testing is performed manually. The proposed method proceeds in four phases. Firstly, screenshots of a Web page are taken on the one hand on a browser-under-test, and on the other hand, on a baseline browser where the correctness of the Web page rendering has been previously asserted. Secondly, image segmentation is applied to the resulting screenshots in order to uniformly split them into regions. Thirdly, computer vision techniques are applied to match and compare regions extracted from the browser-under-test and those extracted from the baseline browser. Finally, machine learning techniques are used to classify the identified differences into acceptable differences vs. incompatibilities, based on sample pairwise comparisons by human testers.

The proposed method is embodied in a commercial tool called Browserbite. The method has been evaluated via experiments involving 140 pages. The evaluation shows that the proposed method achieves an F-score exceeding 90% and outperforms state-of-the-art cross-browser testing tool based on DOM analysis.

This article is a substantially extended version of two previous conference publications [6], [7]. Reference [6] describes the machine learning component of the Browserbite method. Reference [7] is a tool demonstration paper that outlines the high-level architecture and functionality of the Browserbite tool. The present article extends these previous papers by presenting the Browserbite method in an end-to-end manner, including the screenshot capture, image segmentation and region comparison phases.

The rest of the paper is structured as follows. Section 2 presents the screenshot capture phase of the Browserbite method. Next Section 3 introduces the segmentation and comparison phases and presents a first evaluation showing that the proposed segmentation and comparison methods outperform an existing state-of-the-art method for DOM-based cross-browser testing. Section 4 presents how the output of the segmentation and comparison methods can be further refined via machine learning techniques in order to tune the technique based on input provided by users. Finally Section 5 discusses related work while Section 6 draws conclusions and outlines directions for future work.

## 2. SCREENSHOT CAPTURE

This section describes the screenshot capture phase of Browserbite. The input of this phase is the URL of the Web page under test and the browser configurations under which testing is to be performed. The output is one image of the entire Web page rendering for each browser configuration. These images are used for incompatibility detection in later stages.

Rendering Web pages on different configurations is a computationally demanding process, involving the launch of a number of virtual machines, which reproduce each browser configuration (device emulator, operating system, browser and settings). To support a large number of iterative tests, Browserbite has to produce test results in a few seconds or up to minutes. Several optimizations are applied to achieve this goal. Loading and capturing a screenshot of a webpage can take up to tens of seconds. In order to produce testing results in seconds, screen capturing had to be parallelized. This means that virtual machines producing screenshots are run in parallel. Also, to conduct many test requests from a potentially large pool of concurrent users, a queueing system was added using Resque [8] – a Redis-based Ruby library for queuing and backend processing of different tasks. Different workers (Ruby processes) are used for different tasks: screenshot capture, image resizing and segmentation and region comparison as shown in Figure 4. This architectural choice makes the system asynchronous, allowing us to scale the system horizontally.

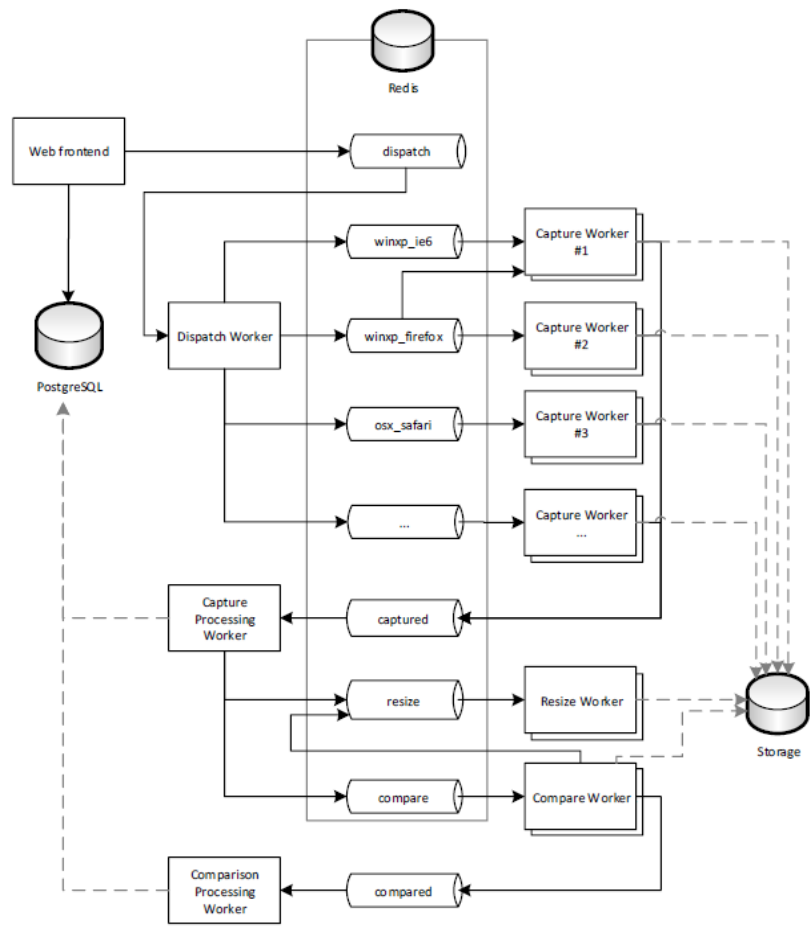

Figure 4. Queues and sequences in Browserbite capture system [9]

The "Capture Worker" is responsible for screenshot generation. This worker is responsible for opening the browser, loading the Web page, configuring the browser viewport and taking the screenshot. A capture woker reserves a job by polling one or several queues that correspond to the installed browsers of the underlying operating system. For each configuration there is a specific crop settings file that includes exact coordinates of the browser viewport. When a worker is first started and when crop settings for a specific configuration are not found, the crop dimensions are calibrated using a test Web page. During this procedure the size and position of the viewport and its scrollbars, as well as some browser-specific attributes, are measured and saved into a configuration file. These parameters are then used for screen capturing.

The capture worker performs browser automation based on Selenium WebDriver [11] software. WebDriver is based on JSON communication between the WebDriver server and client. This solution allowed us to automate most of the browser manipulations, except for the fact that WebDriver library does not have full OSX Safari configuration support, so we used instead the WatiR Ruby library [12] for OSX Safari automation.

Two alternative methods are used by the capture process to acquire a full-page image of a Web document. The first method scrolls through a Web page and stitches image sections into a full size image. The second method resizes a browser's window to match the full size of the Web page document after which the image can be captured at once. The second method is much faster and more reliable compared to the first one. Additionally, using the first method, Web page elements with fixed position on the screen can appear multiple times on different image segments when the page is scrolled – and hence this duplication needs to be detected and eliminated when stitching multiple images together, which may sometimes lead to inaccuracies in the resulting screenshot. Hence the second method is only used when the first method is not supported by a given platform (e.g. OS X together with WatiR does not support window resizing, but the Windows API does).

To decrease Web page loading times, all virtual machines are routed via a proxy server. For this we use the Squid [10] open source web cache and proxy server. This enables us to reduce the load for the remote web server and increase the loading speed. Web page data are cached for a sufficient amount of time so that all shot workers can load the Web page directly from the cache.

3. SEGMENTATION AND COMPARISON

This section describes the core of the Browserbite method, namely segmentation and comparison (cf. Figure 5). The purpose of segmentation is to partition the image (be it the baseline image or the image under test) into smaller comparable Regions-Of-Interest (ROIs). In the second stage, the ROIs extracted from the image of the baseline browser (herein called ROIB) are compared to the ROIs of the image of the browser under test (herein called ROIT). The following subsections outline each of these two steps in turn.

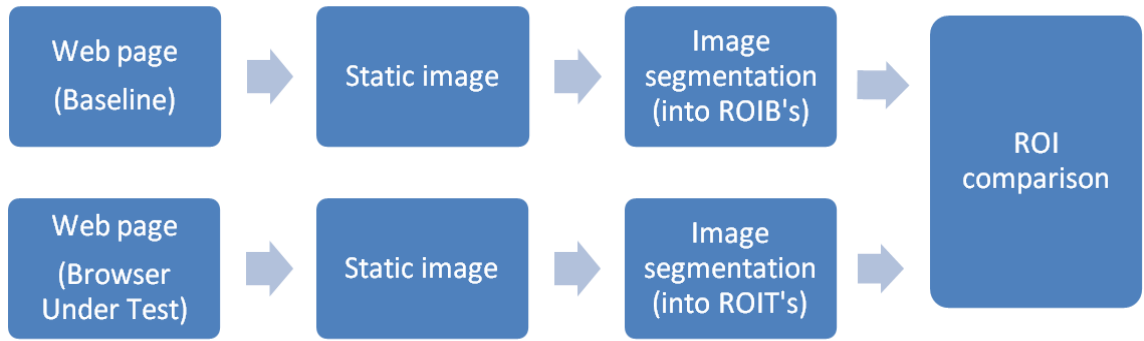

Figure 5. Comparison methods principal layout [13]

### 3.1 *Web page* segmentation

As discussed in Section 1, the Browserbite method is based purely on image processing. In particular, for the purpose of segmentation, the method attempts to mimic the human visual system. It is has been observed that the human visual system focuses more on regions with intensity changes [16]. The latter observation entails that image features related to edges and corners tend to carry most of the hints for visual segmentation. Accordingly, the proposed segmentation technique (cf. Figure. 7) puts emphasis on isolating and processing corner pixels.

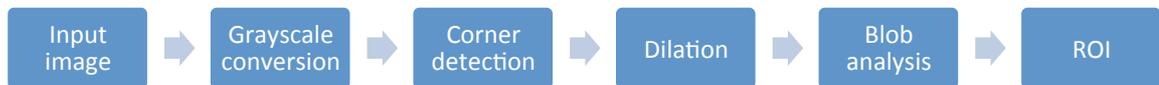

Figure 6. Principal layout of the segmentation technique

As we focus on intensity changes, rather than colour changes, the segmentation method starts by converting the input image into a grayscale image. Next, a corner detection transform is applied to the image to find regions with high frequency details. For this step, we employ the corner detection technique of Harris & Stephens [17]. As a result of this step, a binary image is formed from corner pixels (cf. Figure 7), where corner pixels have a value of one and the background is set to zero.

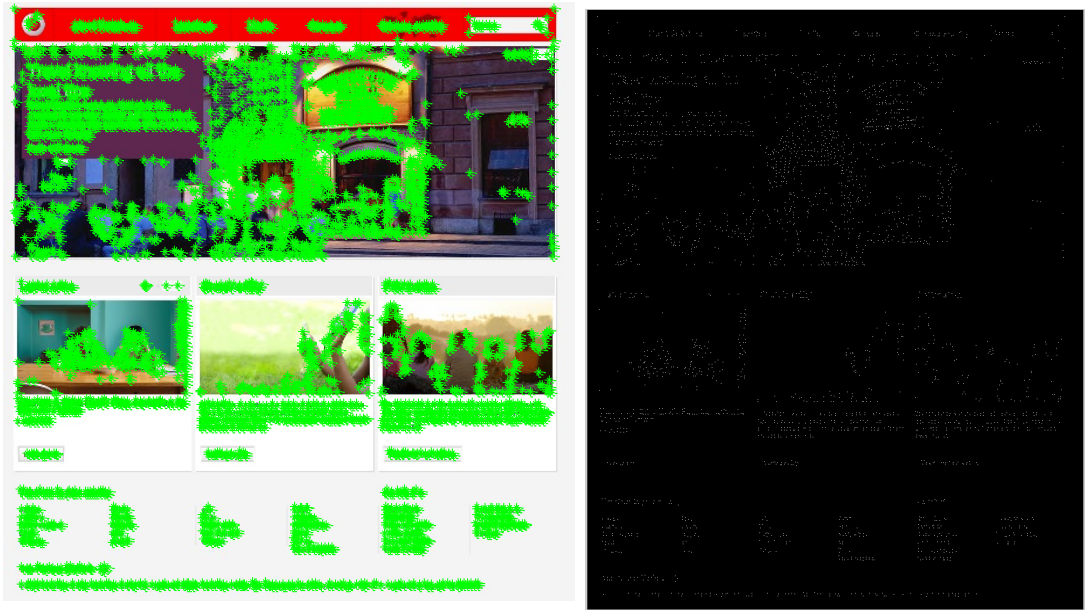

Figure 7. Input image with marked Harris corners (left) and corresponding binary image (right) taken from www.vodafone.com - corner pixels appear as white pixels on the black background

In the next phase, the binary image is processed using vertical and horizontal dilation transforms. This process combines densely situated corner pixels into connected regions. As a result, a binary image with discrete regions is created. By applying blob analysis to this binary image, different blobs are separated into stand-alone ROIs. Figure 8 demonstrates a binary image with separate ROIs with bounding boxes. Each ROI is marked with a unique colour.

The dilation parameter is decremented until the largest ROI side dimension is smaller than 300 pixels or the number of iterations is reached. The extent of the dilation is between 2 - 10 pixels depending on the sizes of the resulting ROIs. Figure 9 shows an input webpage image with the resulting web page segments.

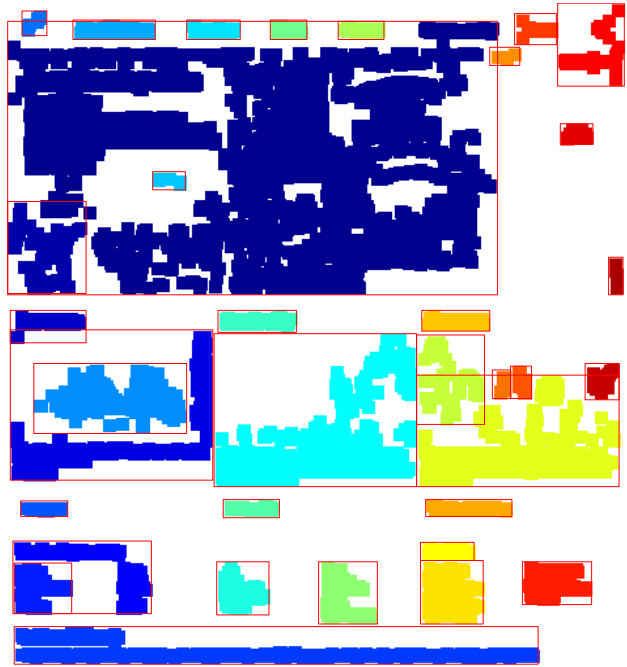

Figure 8. Corner features joined into separate regions

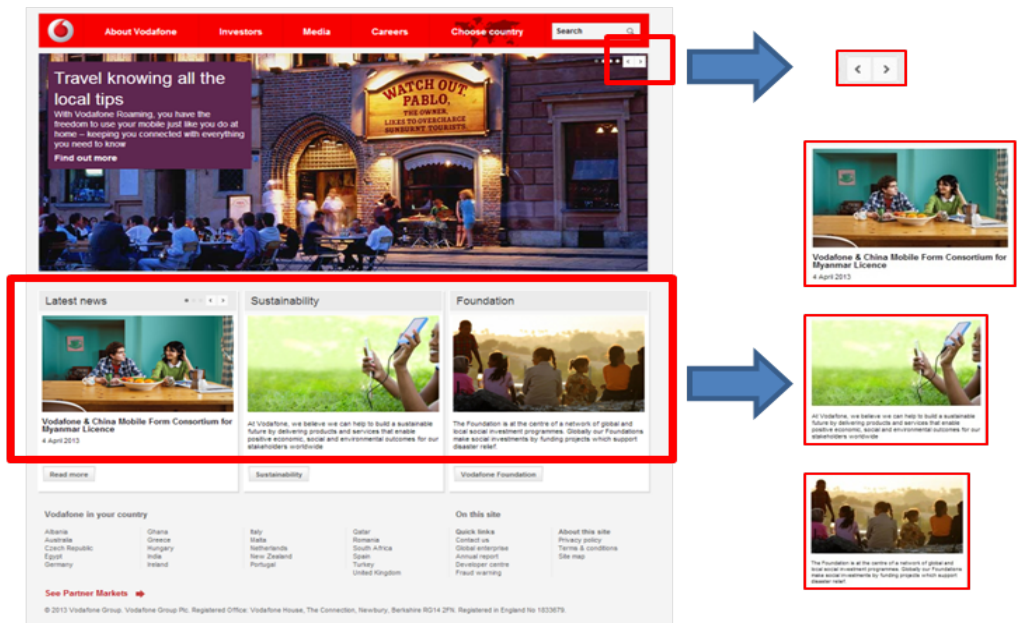

Figure 9. An example Web page with selection of segmented sections (Chrome) (www.vodafone.com)

We compared the proposed Browserbite Web page segmentation technique against the most well known and cited Web page segmentation technique called VIPS [14]. A Vision-based Page Segmentation Algorithm (VIPS) is a method published by Microsoft Research in 2003, which used the DOM data to visually segment Web pages. An example Web page image was extracted from the original VIPS paper. The extracted image had a size of 800 x 1003 pixels. The Browserbite segmentation does not need DOM input, so results can be directly compared. Figure 10 shows the segmentation output from VIPS and Browserbite side by side. The results are similar but not identical. Browserbite produces a greater number of segments, which may simplify the comparison process.

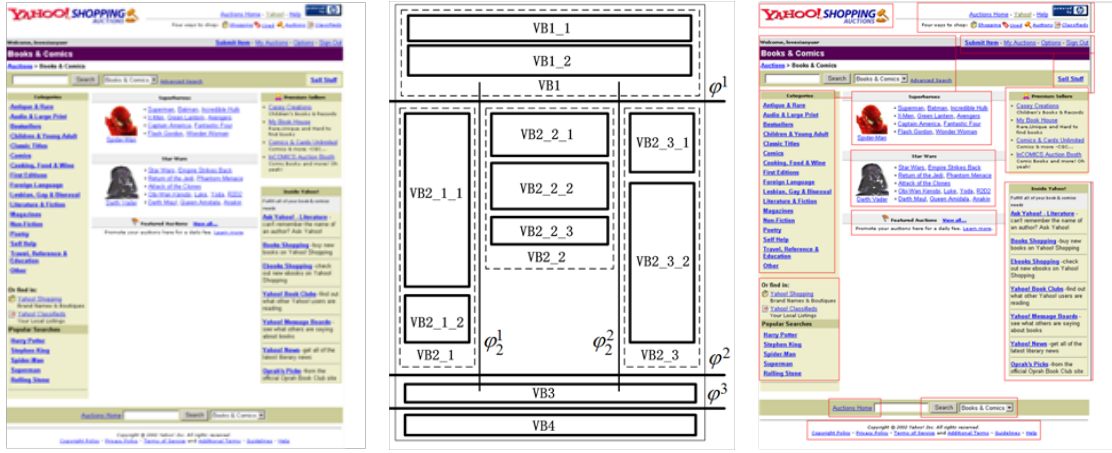

Figure 10. VIPS segmentation against Browserbite segmentation; Original image (left), VIPS segments (middle) and Browserbite segments (right)

*3.2. Image comparison*

As a result of applying segmentation to the baseline image and the image under test, we obtain a number of ROIs (ROIBs for the baseline image, and ROITs for the image under test). The comparison and matching of ROIBs with ROITs is based on feature extraction. For each ROI, the following feature parameters were calculated: size, (x,y) co-ordinates and image raw moments. These parameters were used in the comparison stage to find incompatibilities. Specifically, for the raw image moment calculation, the following equation was used:

$$M_{ij} = \sum_{(i,j) \in W} x^i y^j I(x,y) ,$$

1)

where $M_{ij}$ is the image moment, $i$ and $j$ are the moment orders, $x$ and $y$ are the coordinates [18]. For the centroid co-ordinate calculation, zero and first order moments were divided:

$$\bar{x} = \frac{M_{10}}{M_{00}} ,$$

2)

$$\bar{y} = \frac{M_{01}}{M_{00}}.$$

3)

Additionally, first and second order central moments were calculated as follows:

$$\mu'_{11} = \frac{M_{11}}{M_{00}} - \bar{x}\bar{y},$$

4)

$$\mu'_{20} = \frac{M_{20}}{M_{00}} - \bar{x}^2,$$

5)

$$\mu'_{02} = \frac{M_{02}}{M_{00}} - \bar{y}^2.$$

6)

First and second order central moments were used for the ROI orientation $\theta$ calculation, using the following formula [19]:

$$\theta = \frac{1}{2}\tan^{-1}\left(\frac{2\mu'_{11}}{\mu'_{20} - \mu'_{02}}\right).$$

7)

Image orientation is used as a ROI geometrical property. It is less sensitive to dithering noise and to small geometrical changes. In the comparison stage all ROIs are compared pairwise in order to find matching combinations between the ROIB and ROIT. The following ROI parameters are used for matching: ROI centroid position on the baseline image, ROI size and ROI image orientation. If a matching pair is found, the ROIT is considered to be compatible with ROIB.

If a matching pair has not been found, the ROIB is additionally cross-correlated with a ROI search region to find co-ordinates with the highest correlation index. For correlation comparison, the sum of squared differences is calculated:

$$SSD = \sum_{(i,j) \in W} [I_1(i,j) - I_2(x+i, y+j)]^2 , \qquad 8)$$

where $SSD$ is the sum of squared differences, $I_2$ is the ROI image, $I_1$ is the search region, $x$ and $y$ are the coordinates of the ROI on the search region [20]. Search region is defined by the following equation:

$$w_{SR} = w_{ROI} + d, \qquad 9)$$

$$h_{SR} = h_{ROI} + d, \qquad 10)$$

where $w_{SR}$ is the search region width, $w_{ROI}$ is the ROI width, $h_{SR}$ is the search region height, $h_{ROI}$ is the ROI height and $d$ is the search region size tolerance.

If the SSD index exceeds a threshold, the ROI pair is declared incompatible. This means that inside of the search region there are no similar visual elements. The threshold value was determined empirically over 100 sample cases.

If one or more mismatching ROI pairs are found when comparing the image of a browser-under-test and that of a baseline browser, the configuration of the browser-under-test is declared incompatible.

### 3.3. Initial Evaluation

Below we present an evaluation of the accuracy of the "bare bones" Browserbite method based purely on image segmentation and comparison (i.e. without the classification phase). For this evaluation, we selected the 140 most popular Estonian Web pages (.ee domain) according to *Alexa.com* [21]. Each Web page was rendered on three web browsers on the Windows 7 operating system: Google Chrome (used as baseline browser), and Internet Explorer 8 and Firefox 16.0.1 (used as browsers under test). For each Web page, the screenshots taken on each browser under test were visually compared against the corresponding screenshot on the reference browser. This visual testing was conducted by the authors of this paper. To reduce bias, we applied strictly a pre-defined set of defined criteria to determine what constitutes an incompatibility:

- Visibility differences: if any element in a page is visible in the baseline browser, but not visible in a browser under test, then an incompatibility is declared between the browser under test and the baseline browser for the page in question;

- Position and size differences:

    o The position of an element may not differ by more than 40 pixels along the vertical or horizontal axes, otherwise an incompatibility is declared;

    o The size difference (height or/and width) of the same element rendered in different browsers may not exceed 15 pixels, otherwise an incompatibility is declared.

- Appearance differences:

    o The element's colour is not visibly different, otherwise an incompatibility is declared;

    o The element's font, font style (bold, italic, underlined) and font size are the same in two screenshots, otherwise an incompatibility is declared;

    o The content of every element must be the same in the screenshots, otherwise an incompatibility is declared.

The same pairs of Web pages were compared using bare-bones Browserbite (i.e. Browserbite without the classification phase). To have a state-of-the-art baseline for comparison, we also gave the pairs of Web pages as input to a commercial tool called Mogotest [4], which implements a DOM-based compatibility testing method. Using the incompatibilites found during manual testing, we measured the performance of the Browserbite method in terms of precision, recall and F-score with their standard definitions.

Table 1. Table of results

|  | Precision | Recall | F-score |
|---|---|---|---|
| *Manual* | - | - | - |
| *Bare-bones Browserbite* | 0.66 | 0.98 | 0.79 |
| *Mogotest* | 0.75 | 0.82 | 0.78 |

The evaluation results are given in Table 1. The results show that bare bones Browserbite has a very high recall (only 2% of false negatives), but a lower precision compared to Mogotest – for an overall similar F-score. The low precision of bare-bones Browserbite motivated us to add a classification phase to the method, as outlined in the following section.

## 4. CLASSIFICATION

The aim of the classification phase is to reduce the number false positives during the pairwise comparison of ROIs discussed in the previous section. We explore two variants of this problem:

• Binary classification: In this variant the aim is to classify each potential incompatibility reported by bare-bones Browserbite into two categories: true positive (the potential incompatibility is perceived as such by a user) and false positive (the potential incompatibility is not perceived as such by a user).

• Quaternary classification: In this variant we aim to classify potential incompatibilities into four finer-grained categories as discussed below.

The rest of the section presents the datasets used for training/testing the models, followed by descriptions of the employed features and machine learning techniques.

*4.1. Dataset and setup*

To train the classifiers, we start from the same dataset described in Section 3.3 and use the same browser-under-tests and baseline browser configurations. The 140 Web pages are given as input to bare-bones Browserbite. We extract all detected incompatibilities at the level of ROIs. At this level, each incompatibility reported by (bare-bones) Browserbite takes the form of a pair consisting of an ROIB (baseline ROI) and an ROIT (ROI of the browser-under-test).

Browserbite found about 20 000 potential incompatibilities. This is substantially more than the number of page-level incompatibilities reported in Section 3.3, because one incompatibility in a pair of Web page renderings typically shows up as many incompatibilities at the level of ROI pairs.

### 4.1.1 Golden standard for binary classification

In order to construct a standard for training and testing the binary classification model, the first author of this paper manually classified each ROI into positive sample (i.e. correct result returned by Browserbite) or negative sample (i.e. incorrect result). In this manual classification, a potential incompatibility was deemed to be an actual incompatibility (i.e. put in the "Correct" class) if there is a major layout or formatting difference. In other words, the positioning of a fragment of text or a visual element in the ROIB is clearly different from that in the ROIT or vice-versa, or the formatting or colors in one image are distinct from those in the other. Otherwise the sample was labeled as "Incorrect".

Given the high number of ROIs, only a subset could be manually classified with reasonable effort. Moreover, we aimed at obtaining a balanced set of samples (50% true positives and 50% false positives) so as to avoid affecting the classification models by skewing them towards one class. Thus, ROIs were manually classified in random order until 1200 positive samples were found. Past this point, we did not retain any further positive samples but continued sampling and manual classifying until 1200 negative samples were found, resulting in a balanced set of 2400 samples.

### 4.1.2 Golden standard for quaternary classification

In practice, the notion of incompatibility is not clear-cut. Some incompatibilities are rather minor and do not affect in any significant way the functionality or structure of a page, while others do. Accordingly we also attempted to construct quaternary classification models according to the following class definitions:

- Class *C1*: there is definitely no difference and thus (ROIB, ROIT) pair cannot be considered to be an incompatibility;

- Class *C2*: there is minor layout or formatting difference, but it is not significant, so the (ROIB, ROIT) pair cannot be considered to be an incompatibility;

- Class *C3*: there is a major layout or formatting difference (cf. example in Figure 11);
- Class *C4*: there is a critical difference that affects navigation or other functionality of the page (e.g. a missing button) or that affects its aesthetics in a significant manner (see Figure 12).

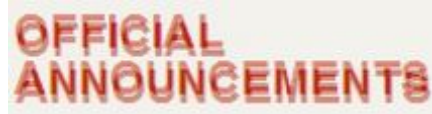

Figure 11. Non-significant layout difference (www.rik.ee)

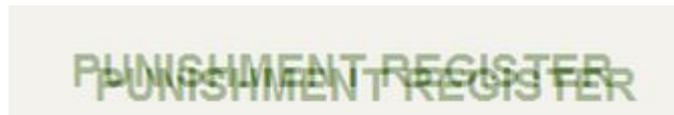

Figure 12. Significant layout difference (www.rik.ee)

Quaternary classification clearly involves a certain level subjectivity, as for example the difference between critical and non-critical can arguably differ across subjects. Accordingly, for the construction and evaluation of quaternary classification models, we recruited 40 subjects through social media and asked them to classify the above set of ROIs into the four mentioned classes. Respondents were University students in the range of 20-25 years from 6 countries (Estonia, Russia, Ukraine, Germany, Italy and Hungary). The subjects came from different specialties: 60% with the IT background, 20% with economics and business background, 10% with philological background, and 10% others.

Each respondent classified between 200 and 400 (ROIB, ROIT) pairs randomly sampled from the dataset with replacement. As a result we obtained between 8 and 15 classifications per pair. To achieve uniformity, we randomly trimmed the dataset so that each (ROIB, ROIT) had only 8 classifications (i.e. 8 subjects per pair). Additionally, the first author of the paper classified each of the pair using the four categories, so that one the end we obtained 9 classifications per (ROIB, ROIT).

The inter-rater reliability of the resulting dataset is 0.94[1], which indicates little disagreement between judges. Accordingly, we calculated the average class for each ROI (rounded to the nearest integer) and used the resulting "average class" as source of truth. The resulting distribution of samples across the four classes is shown in Figure 13.

---

[1] Calculated using the online Inter-Rater Reliability Calculator at http://www.med-ed-online.org/rating/reliability.html

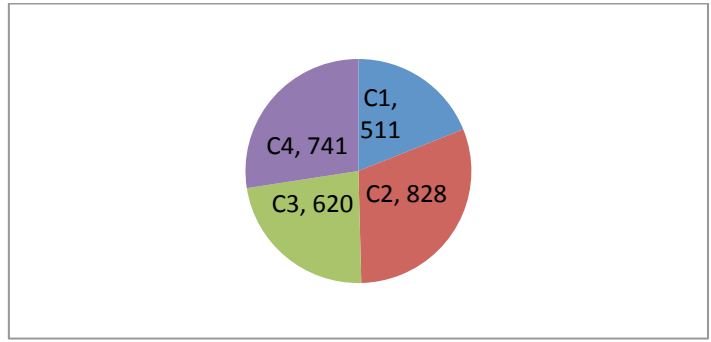

Figure 13. Dataset distribution between four classes

*4.2. Features*

We recall that an incompatibility reported by Browserbite consists of a pair (ROIB, ROIT) where ROIB is an ROI in the baseline image and ROIT is a corresponding ROI in the IUT. In case of a missing or additional ROI, ROIT and ROIB can take null values.

Given a pair (ROIB, ROIT), we extract, the following 17 features to build a sample for constructing classification models:

- 10 *histogram bins* (h0, h1, … h9). These 10 integers encode the image histogram of the ROIB. 10 discrete bins represent pixel intensity distribution across the entire ROI image;

- Correlation between the ROI in the baseline image and ROI in the Image of the browser Under Test (IUT). This is a number between zero and one. It is zero in case of low correlation between ROIB and ROIT;

- Horizontal and vertical position of the ROIB (X and Y coordinates) on the baseline image;

- Horizontal and vertical size of ROIB (width and height) of the baseline image;

- Configuration index – a numerical identifier of the browser-platform combination of the IUT. Browserbite supports 14 browser-platforms combinations, thus this is an integer between 1 and 14;

- Mismatch Density MD = E / T, where E is the number of ROIs in the IUT that are not matched 100% to an ROI in the baseline image, and T is the total number of ROIs in the IUT. Note that this is a feature of the IUT itself rather than a feature of an ROI inside the IUT. However, for the sake of convenience when constructing the machine learning models, we make the MD a feature of each ROI. All ROIs extracted from the same IUT will have the same MD (the MD of their enclosing IUT).

*4.3. Classification methods*

The possible decision for the described problem is to build a classification model, which is able to predict whether the incompatibility is identified correctly on basis of the parameters of each ROI. We selected two well-known classification methods: decision trees and artificial neural networks [23]. Specifically, we used the implementation of these machine learning techniques provided in Matlab.

The use of decision trees in this context is motivated by the fact that it provides a convenient way to interpret the results. By analyzing the decision tree, we can obtain insights into the thresholds and determine whether a given potential incompatibility reported by Browserbite is an actual incompatibility or not.

Neural networks imitate the brain's ability to sort out patterns and learn from trials and errors, discerning and extracting the relationships that underlie the data with which it is presented. A key advantage of neural networks is their ability to adjust themselves to the data without any explicit specification of functional or distributional form.

We selected the three-layered feed-forward neural network for the current problem. The first layer (the input layer) consists of 17 neurons that correspond to the number of features. The output layer has 2 neurons for binary classification and 4 neurons for quaternary. As the dataset is not linearly separable there is a need to introduce one or more additional "hidden" layers. In practice, adding a second hidden layer can solve few problems that cannot be solved with a single hidden layer. This general rule was confirmed by our experiments, which showed that adding two hidden layers did not improve the accuracy (F-score) with respect to a single hidden layer. The results reported below are with one hidden layer.

The number of neurons in the neural network is another important parameter, as too few hidden neurons can cause underfitting so that the neural network cannot learn the details. Conversely, a too large number of hidden neurons can cause overfitting, as the neural network starts to learn insignificant details. Accordingly, the number of hidden neurons was determined experimentally as discussed below.

## 4.4. Classifier evaluation

The trained and tested classification models using both decision trees and neural networks based on the previously presented set of features. Classification accuracy was measured in terms of precision, recall and F-score. The robustness of the results was validated using a five-fold cross-validation method. In other words, the dataset was partitioned into five equal parts, four parts were used to train a model and the remaining one was used to test the model. This process was repeated 5 times with each part playing the testing role once. The results from each fold were then averaged to produce a single measurement of precision, recall and F-score for each method (decision tree and neural network). We also measured execution time for creating the classifiers using the "tic/toc" function in Matlab.

The results for the binary and quaternary cases are given in Table 2 and Table 3 respectively. In the binary case, it can be seen that neural networks outperform clearly decision trees. In the quaternary case, we observe again that neural networks outperform decision trees across all classes, but less markedly so. Execution time for both methods is comparable.

Table 2. Binary classification results

| Measurement | *Decision tree* | *Neural Network* |
|---|---|---|
| Precision | 0.85 | 0.963 |
| Recall | 0.78 | 0.887 |
| F-score | 0.81 | 0.923 |
| Exec. Time (sec.) / ROI pair | 1.43 | 1.71 |

Table 3. Quaternary classification results

| Measure | *Decision tree* | | | | *Neural Network* | | | |
|---|---|---|---|---|---|---|---|---|
| | *C1* | *C2* | *C3* | *C4* | *C1* | *C2* | *C3* | *C4* |
| Precision | 0.56 | 0.53 | 0.65 | 0.61 | 0.65 | 0.69 | 0.69 | 0.77 |
| Recall | 0.61 | 0.67 | 0.53 | 0.59 | 0.52 | 0.75 | 0.64 | 0.73 |
| F-score | 0.58 | 0.59 | 0.58 | 0.6 | 0.58 | 0.72 | 0.66 | 0.75 |

| Exec. Time (sec.) | 1.72 | 1.97 |
|---|---|---|

In order to determine the appropriate number of hidden neurons we trained the neural network with different number of neurons and calculated the F-score for each trained model (cf. Figure 14 both for the binary and the quaternary classification tasks). We experimentally found that the peak in F-score is reached for a number of hidden neurons of 11, both for binary and for quaternary classification cases. The results reported in Table 2 and Table 3 are for the neural network with 11 neurons.

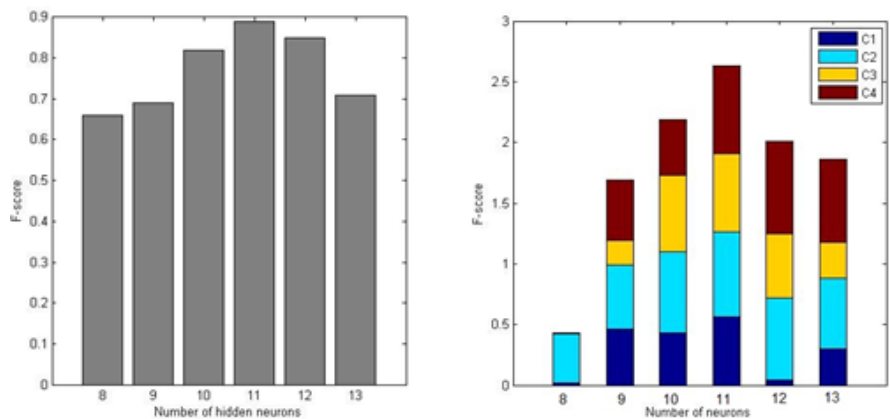

Figure 14. F-score for binary and quaternary classification

*4.5. Decision tree example*

In order to understand which features play a role in the detection of false positives, we present the decision tree for the binary classification problem. The decision tree consists of 393 nodes. However, it is possible to perform a tree pruning by merging leaves on the same tree branch. We used the pruning sequence calculated by Matlab by default when building this decision tree. The first five levels of pruned decision tree are presented in Figure 15.

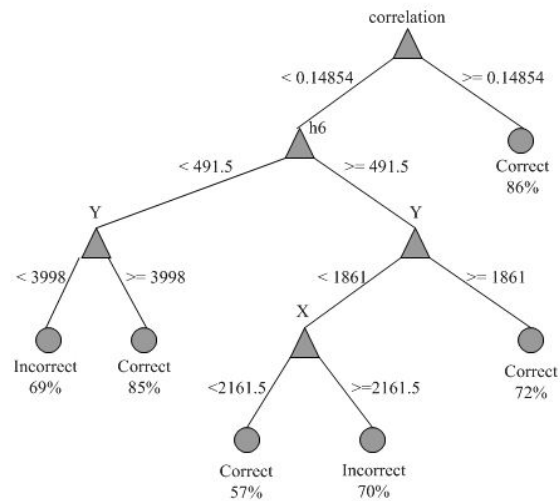

Figure 15. Sample pruned decision tree for the binary classification

According to the pruned decision tree, when the correlation index of the (ROIB, ROIT) pair is greater than 0.14854, it will be classified as an actual incompatibility with probability of 86%. This result is expected as this index captures to the differences between a pair of images. Less expected is the fact that Y > 3998 is associated with true positives, meaning that incompatibilities detected at the bottom of a page are more likely to be true incompatibilities compared to incompatibilities detected at the top of the page. This may be due to the fact that small displacements of elements at the top of the page ripple down into more visible incompatibilities at the bottom of the page, and thus clear incompatibilities are more likely to occur at the bottom of the page.

## 5. RELATED WORK

Broadly speaking, there are two complementary families of methods for automatic cross-browser incompatibility testing: *trace-level testing* and a *page-level testing*. In trace-level testing, different states and state transitions of a Web page are compared. From each configuration, a state tree is extracted and tree branches between a baseline browser and a browser under test are compared. In contrast, this paper is concerned with page-level testing, where the goal is to detect incompatibilities in the rendering of one given Web page across multiple browsers. Accordingly, the following review of related work focuses mainly on page-level testing. Trace-level testing methods are reviewed only insofar as they support page-level testing as well.

As discussed in Section 1, page-level cross-browser testing approaches rely on an analysis of the DOM of a Web page and/or an analysis of the Web page renderings of a Web page across multiple browsers. Browserbite relies purely on image processing.

Two commercial tools Browsera [24] and Mogotest [4] use DOM extraction and comparison for cross-browser testing. As discussed in Section 3.3, we have found that the F-score achieved by bare-bones Browserbite (without the classification module) is comparable to that of Mogotest. Once the classification module is factored in, Browserbite's performance is significantly enhanced, especially when using a neural networks classifier.

Mahajan and Halfond [25] describe a method for discovery of bugs between initial design mockups and actual Web pages. The mockups are used as oracles, which are compared against Web pages using pixel based comparison. Although effective for the described application scenario, it is not suitable for cross browser testing due to pixel-by-pixel comparison limitations.

Mesbah and Prasad [26] proposed a trace-level cross browser testing method. They have divided their solution into two main stages: web application crawling (using different configurations) and model comparison. By crawling a web application, a navigation model of the application is constructed. The navigation model is a state tree in which nodes correspond to DOM trees, and edges represent state transitions. For cross-browser testing, two types of comparisons are conducted: trace-level and page-level. In contrast to Browserbite, this technique focuses exclusively on DOM-level differences, ignoring the final visual rendering result, which poses several issues as illustrated in Section 1.

A tool called WebMate [27]–[29] uses Selenium [11] to extract a state tree of a Web page. This state tree is compared across browsers to find behavioural differences. Also, layout testing is described in [29]. It is based on Web page element extraction and comparison using the DOM data, but the whole page layout incompatibility testing is not covered.

Choudhary et al [30], [33] have used image processing techniques in combination with DOM to compare Web page segments. They used Earth Movers' Distance to calculate differences between two image segments. Image segments were extracted from the image of a Web page using DOM coordinates. Different DOM parameters across different configurations can cause variations in image segmentation. Different segment sizes complicate image processing based comparison. In our research we used purely image processing based segmentation, which will produce more uniform segments across browsers, independent of the DOM structure and parameters. In addition, WebDiff uses identical viewport size for all configurations, whereas Browserbite can handle different viewport sizes as long as screen resolutions are identical.

CrossCheck [31] combines page-level methods [26] and trace-level methods [30], with a decision tree classifier added to address problems caused by a simple histogram level comparison. A three stage comparison was proposed consisting of a trace-level, DOM and visual phase. From each stage a number of features were extracted. These features were later used in conjunction with a classifier tree. In contrast to Browserbite, CrossCheck uses DOM data as one of the system inputs. Due to DOM data inconsistency across browsers this can lead to false positive results as discussed in Section 1.

X-PERT [32] combines a model collector (crawler) and a model comparator. Although this work includes detailed description of the solution, it is unclear which are the specific criteria for layout incompatibilities are used. Results of the X-PERT prototype were compared against CrossCheck based on tests conducted on 14 Web pages. The precision of X-PERT was 76% and the recall 95% (comparable to the one we recorded for bare-bones Browserbite) but the very low number of Web pages under test in [32] puts into question the statistical significance of the results.

A tool called Applitools Eyes [34] uses pure visual Web page image segmentation and comparison. Their patent applications [35], [36] describe an approach using thresholding and morphological filtering based image segmentation. Pixel-based comparison is proposed for the segment comparison. In contrast, Browserbite employs feature-based image segmentation, which is less sensitive to gradient backgrounds and imperceptible differences.

## 6. CONCLUSION

This paper has presented the end-to-end cross-browser testing method underpinning the Browserbite tool. At present, Browserbite is a commercial SaaS (Software as a Service) tool available at http://*www.browserbite.com*. It supports more than 15 desktop and mobile browser configurations. The system detects incompatibilities at the level of ROIs as discussed in this paper and then aggregates them at the level of the page. Detected incompatibilities are highlighted in red regions and overlaid on top of the corresponding screenshots as illustrated in Figure 16. Browserbite has over 10 000 registered users as of January 2015.

The results of the experimental evaluation reported in this paper, as well as Browserbite's extensive usage in practice, demonstrates the feasibility of high-accuracy cross-browser testing based on image processing.

A limitation of the method presented in this paper is that cross-browser testing is restricted to the level of single static Web pages, such that testing can be performed by comparing pairs of screenshots. The commercial edition of Browserbite additionally supports limited forms of client-side dynamic Web page testing by taking screenshots of a page after a few seconds or after specific event occurrences and checking if the rendering has changed in-between the two screenshots. In the latter case, cross-browser compatibility is checked separately for the two taken screenshots. Browserbite also allows testers to specify the Web page to be tested by means of a sequence of user actions (e.g. hovering, form field editing and clicks) starting from a given Web page. Sequences of user actions are captured in one configuration, saved as a script, and later replayed in other configurations. In this way, testers can use Browserbite to test (individual) pages within a Web page flow.

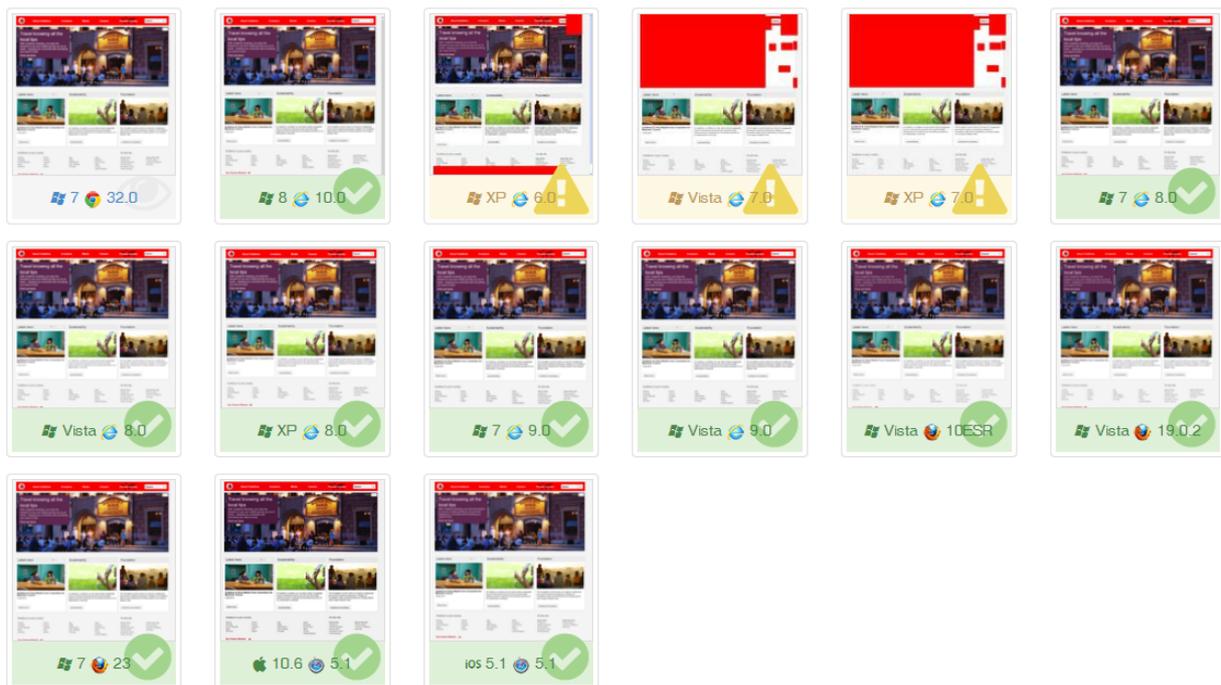

Figure 16. Results of *www.vodafone*.com test

Ongoing development of Browserbite is targeted at enhancing its support for testing Web page flows towards trace-level testing, as well as extending Browserbite to non-traditional Web platforms such as smart TVs, billboards and other devices.

Another avenue for future work is to further study the case of quaternary classification, for which the accuracy achieved by neural networks leaves room for improvement. To this end, we need to consider further features and conduct larger-scale experiments. A related direction is to evaluate the proposed techniques with different types of stakeholders involved in Web application development (e.g. Web designers versus testers versus developers). In this respect, one can hypothesize that classification models for designers would be different than those for developers, for example.